\documentclass{PoS}
\newcommand{\beq}{\begin{equation}}
\newcommand{\eeq}{\end{equation}}
\newcommand{\bea}{\begin{eqnarray}}
\newcommand{\eea}{\end{eqnarray}}
\def\Id{ \mbox{1\hspace{-1.2mm}I} }

\def\BA{\begin{eqnarray}}
\def\EA{\end{eqnarray}}
\def\BAN{\begin{eqnarray*}}
\def\EAN{\end{eqnarray*}}
\def\nn{\nonumber\\}

\def\u{{\bf u}}
\def\d{{\bf d}}
\def\s{{\bf s}}
\def\c{{\bf c}}

\def\q{{\bf q}}
\def\Q{{\bf Q}}
\def\ubar{\bar{\bf u}}
\def\dbar{\bar{\bf d}}
\def\sbar{\bar{\bf s}}
\def\cbar{\bar{\bf c}}
\def\qbar{\bar{\bf q}}

\def\g5{\gamma_5}
\def\g4{\gamma_4}
\def\g3{\gamma_3}
\def\g2{\gamma_2}
\def\g1{\gamma_1}

\title{The spectrum of charmonium-like vector mesons in lattice QCD}

\ShortTitle{The spectrum of charmonium-like vector mesons in lattice QCD}

\author{TWQCD Collaboration: 
   \speaker{Ting-Wai Chiu}$^a$,Tung-Han Hsieh$^b$ \\
\llap{$^a$} Department of Physics and Center for Theoretical Sciences, 
            National Taiwan University, Taipei~10617, Taiwan \\
            E-mail: \email{twchiu@phys.ntu.edu.tw} \\
\llap{$^b$} Research Center for Applied Sciences, Academia Sinica, 
            Taipei 115, Taiwan \\
            E-mail: \email{thhsieh@twcp1.phys.ntu.edu.tw}  
}

\abstract{
We present the first lattice results of the spectrum of exotic vector mesons
extracted from the molecular and diquark-antidiquark operators, with
quark fields $(\c\q\cbar\qbar)$, and
$(\c\s\cbar\qbar)/(\c\q\cbar\sbar)$ respectively,
in lattice QCD with exact chiral symmetry.
Our results suggest that
$ X(3872) $ and $ Y(4260) $ are in the spectrum of QCD,
with $ J^{PC} = 1^{++} $ and $ 1^{--} $ respectively.
Moreover, we obtain the spectrum of heavier exotic mesons with
$(\c\s\cbar\ubar)/(\c\u\cbar\sbar)$, $(\c\s\cbar\dbar)/(\c\d\cbar\sbar)$, 
$(\c\s\cbar\sbar)$, and $(\c\c\cbar\cbar)$, 
as the first theoretical predictions from lattice QCD.}

\FullConference{XXIVth International Symposium on Lattice Field Theory\\
                July 23-28, 2006\\
                Tucson, Arizona, USA}

\begin{document}

\section{Introduction}

Since the discovery of $ D_s(2317) $ \cite{Aubert:2003fg} by BABAR
in April 2003, a series of new heavy mesons
with open-charm and closed-charm have been observed by
Belle, CDF, CLEO, BABAR, and BES. 
Among these new heavy mesons, the most intriguing ones are
the charmonium-like states,
$ X(3872) $ \cite{Choi:2003ue},
$ Y(3940) $ \cite{Abe:2004zs},
$ Y(4260) $ \cite{Aubert:2005rm},
$ Z(3930) $ \cite{Uehara:2005qd},
and $ X(3940) $ \cite{Abe:2005hd}.
Evidently, one can hardly interpret all of them as orbital and/or radial
excitations in the charmonium spectrum.
Thus it is likely that some of them are exotic (non-$q\bar q $) mesons
(e.g., molecule, diquark-antidiquark, and hybrid meson).
Theoretically, the central question is whether
the spectrum of QCD possesses these resonances, with the correct
quantum numbers, masses, and decay widths.
Now the most viable approach to tackle this problem from the first principles 
of QCD is the lattice QCD with exact chiral symmetry 
\cite{Kaplan:1992bt,Narayanan:1995gw,Neuberger:1997fp,Ginsparg:1981bj}.

Recently, we have investigated the mass spectrum of closed-charm
exotic mesons with $ J^{PC} = 1^{--} $
\cite{Chiu:2005ey} and $ 1^{++} $ \cite{Chiu:2006hd} respectively, 
in the framework of lattice QCD with exact chiral symmetry.
By constructing molecular and diquark-antidiquark operators
with quark content $ (\c\q\cbar\qbar) $, we measure the  
time-correlation function of each operator, and fit  
to the usual formula 
\BAN
W \{ e^{-m a t} + e^{-m a (T-t)} \}
\EAN
to extract the mass ($m$) and the 
spectral weight ($W$) of the lowest-lying state overlapping with 
the exotic meson operator. 
From the ratio of the spectral weights of two spatial volumes
($20^3 $ and $ 24^3 $) with the same lattice spacing, 
we can infer whether this state is a resonance 
($ W_{20}/W_{24} \simeq 1 $)
or 2-particle scattering state ($ W_{20}/W_{24} \simeq (24/20)^3 = 1.728 $).
  
Our results suggest that $ Y(4260) $ and $ X(3872) $
are in the spectrum of QCD,
with $ J^{PC} = 1^{--} $ and $ 1^{++} $ respectively,
and both with quark content $ (\c\u\cbar\ubar) $. 
Note that we have been working in the isospin limit ($ m_u = m_d $), 
thus our results \cite{Chiu:2005ey,Chiu:2006hd} cannot exclude the 
possibility of the existence of exotic mesons with 
quark content $ (\c\d\cbar\dbar) $,  
even though we cannot determine their mass differences from those with
$ (\c\u\cbar\ubar) $. Moreover, we also observe heavier exotic mesons
with quark contents $ (\c\s\cbar\sbar) $ and $ (\c\c\cbar\cbar) $,
for $ J^{PC} = 1^{++} $ \cite{Chiu:2006hd}, and $ 1^{--} $
\cite{Chiu:2005ey} respectively. 

Now if the spectrum of QCD does possess exotic mesons with quark
content $ (\c\q\cbar\qbar) $, then it is likely that there
are also exotic mesons with other quark contents,
e.g., $ (\c\q\cbar\sbar) $ and $ (\c\s\cbar\qbar) $.
However, in general, whether any combination of two quarks
and two antiquarks can emerge as a hadronic state relies on
the nonperturbative dynamics between these four quarks.
Recently, we have investigated the lowest-lying mass spectrum of  
molecular and diquark-antidiquark operators with 
quark content $ (\c\s\cbar\qbar) $/$ (\c\q\cbar\sbar) $ 
and $ J^P = 1^+ $ \cite{Chiu:2006us}.
Our results suggest that there exists a $ J^P = 1^+ $ resonance around 
$ 4010 \pm 50 $ MeV. It is interesting to see whether this state will be 
observed by high energy experiments.

\section{Lattice setup}

To implement exact chiral symmetry on the lattice, 
we consider the optimal domain-wall fermion proposed by Chiu \cite{Chiu:2002ir}.
From the generating functional for $n$-point Green's function of the quark
fields, the valence quark propagator in background gauge field can be derived
as \cite{Chiu:2002ir}
\bea
\label{eq:quark_prop}
\langle q(x) \bar q(y) \rangle 
= (D_c + m_q)^{-1}_{x,y}, \hspace{4mm} 
D_c = 2 m_0 \frac{1 + \gamma_5 S(H_w)}{1-\gamma_5 S(H_w)}
\eea
where $m_q$ is the bare quark mass, 
      $m_0$ is a parameter in the range $ (0,2) $, 
      $S(H_w)$ is equal to the Zolotarev approximation 
               of the sign function of $ H_w $ 
($H_w = \gamma_5 D_w$, 
 and $D_w$ is the standard Wilson Dirac operator minus $m_0$).
In the limit $ N_s \to \infty $ 
(where $ N_s + 2 $ is the number of sites in the 5-th dimension), 
$ D_c $ is exactly chirally symmetric, i.e., 
$ D_c \gamma_5 + \gamma_5 D_c = 0 $. 
In the continuum limit, the valence quark propagator $ (D_c + m_q)^{-1} \to 
[ \gamma_\mu (\partial_\mu + i g A_\mu) + m_q ]^{-1} $. 
A salient feature of the valence quark propagator (\ref{eq:quark_prop})
is that the bare quark mass $ m_q $ is well-defined for any gauge field 
configuration, unlike the Wilson quark propagator.  
In practice, we compute the valence quark propagator 
by the nested conjugate gradient
\bea
\label{eq:outer_CG}
D(m_q) Z &=& \{ m_q+(m_0-m_q/2) [1+\gamma_5 S(H_w)] \} Z = \Id   \\
\label{eq:inner_CG}
S(H_w) Z &=& \sum_l b_l z_l, \hspace{4mm} (H_w^2 + c_l) z_l = Z 
\eea
where $ b_l $ and $ c_l $ are Zolotarev coefficients 
\cite{vandenEshof:2001hp,Chiu:2002eh}.
To attain the maximal efficiency of our system 
(CPU time and memory usage vs. the precision of the sign function $S(H_w)$), 
we use Neuberger's 2-pass algorithm \cite{Neuberger:1998jk,Chiu:2003ub}
in the inner conjugate gradient (\ref{eq:inner_CG}).

We generate 100 gauge configurations with single plaquette gauge action
at $ \beta = 6.1 $, for two lattice volumes $ 24^3 \times 48 $ 
and $ 20^3 \times 40 $, with the same lattice spacing. 
Fixing $ m_0 = 1.3 $, we project out 16 low-lying eigenmodes of
$ |H_w| $ and perform the nested conjugate gradient in the complement
of the vector space spanned by these eigenmodes. For
$ N_s = 128 $,
the weights $ \{ \omega_s \} $ are fixed with $ \lambda_{min} = 0.18 $
and $ \lambda_{max} = 6.3 $,
where $ \lambda_{min} \le \lambda(|H_w|) \le \lambda_{max} $
for all gauge configurations.
For each configuration, point-to-point quark propagators are computed
for 30 bare quark masses in the range $ 0.03 \le m_q a \le 0.8 $,
with stopping criteria $ 10^{-11} $ and $ 2 \times 10^{-12} $
for the outer and inner conjugate gradient loops respectively.
Then the norm of the residual vector of each column of the quark propagator
is less than $ 2 \times 10^{-11} $
\BAN
|| (D_c + m_q ) Y - \Id || < 2 \times 10^{-11}, \hspace{4mm}
Y = \{1- m_q/(2m_0)\}^{-1}\{Z - 1/(2m_0)\}, 
\EAN
and the chiral symmetry breaking due to finite $ N_s $ is
less than $ 10^{-14} $,
\BAN
\sigma = \left| \frac{Z^{\dagger} S^2 Z}{Z^{\dagger} Z} - 1 \right|
< 10^{-14},
\EAN
for every iteration of the nested conjugate gradient.

Then we measure the time-correlation functions of pseudoscalar and vector meson
operators. We determine the inverse lattice spacing $ a^{-1} = 2.237(76) $ GeV
from the pion time-correlation function, with the experimental input
of pion decay constant $ f_\pi = 131 $ MeV.
The strange quark bare mass $ m_s a = 0.08 $
and the charm quark bare mass $ m_c a = 0.80 $ are fixed
such that the corresponding masses
extracted from the vector meson correlation
function agree with $ \phi(1020) $ and $ J/\psi (3097) $ respectively
\cite{Chiu:2005zc,Chiu:2005ue}.

\section{Four-quark meson operators}

In QCD, there is no reason why any hadronic state composed of 
two quarks and two antiquarks cannot emerge as a resonance.
Thus, it is interesting to see whether such four-quark mesons 
can exist in the spectrum of QCD, and also to identify them 
with any meson states observed in high energy experiments. 
To tackle this problem in the framework of lattice QCD, 
one needs to construct interpolating operators 
which can have good overlap with the 4-quark meson states. 
Otherwise, their signals may not be unambiguously identified,  
due to the limitations in statistics, the lattice size,   
as well as the unphysically-heavy $ \u(\d) $ quark masses 
which have to be chirally extrapolated to their physical values.
Further, for any lattice calculations in quenched approximation, 
one should avoid any operators (e.g., scalar operator $ \cbar \q $) 
which have strong quenched artifacts as $ m_q \to m_{u,d} $.

In general, there are two ways to construct 4-quark meson operators:  
(i) molecular operator, and (ii) diquark-antidiquark operator. 
In the following, we only present the 4-quark operators which have 
good overlap (except $ Y_4 $) with the hadronic states 
in our investigations \cite{Chiu:2005ey,Chiu:2006hd,Chiu:2006us}.
Explicitly, they are
\bea
& J^{PC} = 1^{--}: & \nn
&& M_3(x) = \frac{1}{\sqrt{2}} \left\{
               (\qbar \gamma_5\gamma_i \c)_x (\cbar\gamma_5 \q)_x
              -(\cbar \gamma_5\gamma_i \q)_x (\qbar\gamma_5 \c)_x \right\} \\
&& Y_4(x) = \frac{1}{\sqrt{2}} \left\{
   (\q^T C \gamma_5 \gamma_i \c)_{ax}(\qbar C \gamma_5 \cbar^T)_{ax}
  +(\qbar C \gamma_5 \gamma_i^T \cbar^T)_{ax}(\q^T C \gamma_5 \c)_{ax} \right\}
\\
& J^{PC} = 1^{++}: &  \nn
&& M_1(x) =
\frac{1}{\sqrt{2}} \left\{ (\qbar\gamma_i \c)_x (\cbar \gamma_5 \q)_x
                         -(\cbar\gamma_i \q)_x (\qbar \gamma_5 \c)_x \right\} \\
&& X_4(x) = \frac{1}{\sqrt{2}} \left\{
   (\q^T C \gamma_i \c)_{ax} (\qbar C \gamma_5 \cbar^T)_{ax}
  -(\qbar C \gamma_i^T \cbar^T)_{ax} (\q^T C \gamma_5 \c)_{ax} \right\}
\\
& J^P = 1^{+}: & \nn
&& M_s(x) =
\frac{1}{\sqrt{2}} \left\{ (\qbar \gamma_i \c)_x (\cbar \gamma_5 \s)_x
                          -(\cbar \gamma_i \s)_x (\qbar \gamma_5 \c)_x \right\}
\\
&& D_4(x) = \frac{1}{\sqrt{2}} \left\{
   (\q^T C \gamma_i \c)_{ax} (\sbar C \gamma_5 \cbar^T)_{ax}
  -(\sbar C \gamma_i^T \cbar^T)_{ax} (\q^T C \gamma_5 \c)_{ax} \right\}
\eea
where  
$$
(\qbar \Gamma \Q)_x = 
\qbar_{a \alpha x} \Gamma_{\alpha\beta} \Q_{a \beta x}
$$
denotes a meson operator with antiquark field $ \qbar $ coupling 
to quark field $ \Q $ through the Dirac matrix $ \Gamma $. Here    
$ x $, $ \{ a,b,c \} $ and $ \{ \alpha, \beta \} $
denote the lattice site, color, and Dirac indices respectively. 
The diquark operator is denoted by 
$$
(\q^T \Gamma \Q)_{ax} = 
\epsilon_{abc} \q_{b \alpha x} \Gamma_{\alpha\beta} \Q_{ c \beta x}
$$
where 
$ \epsilon_{abc} $ is the completely antisymmetric tensor, 
$ C $ is the charge conjugation operator satisfying
$ C \gamma_\mu C^{-1} = -\gamma_\mu^T $ and
$ (C \gamma_5)^T=-C\gamma_5 $.    
Thus the diquark transforms like color anti-triplet.
For $ \Gamma = C \gamma_5 $, it transforms like $ J^P = 0^{+} $,
while for $ \Gamma = C \gamma_i $ ($i=1,2,3 $), it
transforms like $ 1^{+} $.

\section{Results and discussions}

In Table \ref{tab:mass_summary}, we summarize our results of 
the masses of the lowest-lying states of the 4-quark operators 
obtained in Refs. \cite{Chiu:2005ey,Chiu:2006hd,Chiu:2006us}, 
where in each case, the first error is statistical, and
the second one is our estimate of combined systematic uncertainty
including:
(i) possible plateaus (fit ranges) with $\chi^2 $/d.o.f. < 1;
(ii) the uncertainties in the strange quark mass and the charm quark mass;
(iii) chiral extrapolation (for the entries containing u/d quarks); and
(iv) finite size effects (by comparing results of two lattice sizes).
Note that we cannot estimate the discretization error
since we have been working with one lattice spacing.

\subsection{$ J^{PC} = 1^{--} $ states} 
 
The molecular operator 
$ M_3 \sim \{ (\qbar \gamma_5 \gamma_i \c)(\cbar \gamma_5 \q)
    -(\cbar \gamma_5 \gamma_i \q)(\qbar \gamma_5 \c) \} $
detects a resonance with mass $ 4238(31)(57) $ MeV
in the limit $ m_q \to m_{u,d} $, 
which is naturally identifed with $ Y(4260) $. 
This suggests that $ Y(4260) $ is indeed 
in the spectrum of QCD, with quark content ($\c\u\cbar\ubar$)
and $ J^{PC} = 1^{--} $.

For the diquark-antidiquark operator
$ Y_4 \sim \{[\q^T C\gamma_5\gamma_i \c][\qbar C\gamma_5\cbar^T]
   +[\q^T C\gamma_5 \c][\qbar C\gamma_5\gamma_i^T \cbar^T] \} $,
it also detects a state with mass $ 4267(68)(83) $ MeV,   
in the limit $ m_q \to m_u $. 
We suspect that this state might be the same resonance captured by
the molecular operator $ M_3 $.
However, we are not sure that this state is a resonance since
the ratio of spectral weights ($ R = W_{20} / W_{24} $)
for two different lattice volumes with the same lattice spacing
deviates from one (the criterion for a resonance)
with large errors as $ m_q \to m_u $.
It is plausible that such a deviation is due to the
quenched artifacts which can be evaded if one incorporates
internal quark loops.

Now, in the quenched approximation, our results
suggest that $ Y(4260) $ has a better overlap with the molecular
operator $ M_3 $ than the diquark-antidiquark operator $ Y_4 $. 
Whether this implies that $ Y(4260) $ behaves more likely as a
$ D_1 \bar D $ molecule than a diquark-antidiquark
meson is subjected to further investigations, especially those
incorporating dynamical quarks.

For molecular and diquark-antidiquark operators with
quark fields ($\c\s\cbar\sbar$), they both detect a resonance around
$ 4450 \pm 100 $ MeV, and for the molecular operator with
($\c\c\cbar\cbar$), it detects a resonance around $ 6400 \pm 50 $ MeV.
These serve as predictions from lattice QCD.

\subsection{$ J^{PC} = 1^{++} $ states}

Both the molecular operator $ M_1 $
and the diquark-antidiquark operator $ X_4 $
detect a $ 1^{++} $ resonance around $ 3890 \pm 30 $ MeV
in the limit $ m_q \to m_{u,d} $, which is naturally identified
with $ X(3872) $.
This shows that $ X(3872) $ is indeed in the spectrum of QCD,
with quark content ($\c\u\cbar\ubar$), and $ J^{PC} = 1^{++} $.
It is interesting to see that  
$ X(3872) $ has good overlap with the molecular
operator $ M_1 $ as well as the diquark-antidiquark operator $ X_4 $.
This is in contrast to the case of $ Y(4260) $, in which $ Y(4260) $ 
seems to have better overlap with the molecular operator  
$ M_3 $ than any diquark-antidiquark operators.
This seems to suggest that $ X(3872) $ is more tightly bounded than
$ Y(4260) $. It would be interesting to see whether this
picture persists even for unquenched QCD.

For $ m_q = m_s $,
heavier exotic meson resonance with $ J^{PC} = 1^{++} $ is also detected,
with quark content ($\c\s\cbar\sbar$) around $ 4100 \pm 50 $ MeV.
This serves as a prediction from lattice QCD.

\subsection{$ J^{PC} = 1^{+} $ states}

Both the molecular operator $ M_s $ and the diquark-antidiquark 
operator $ D_4 $ detect a $ 1^{+} $ resonance around 
$ 4010 \pm 50 $ MeV in the limit $ m_q \to m_{u,d} $. 
Since its mass is just slightly above Y(3940), 
high energy experiments should be able to
see whether such a resonance, say, $ X_s $, exists in some decay channels,
e.g., $ X_s \to K \pi J/\Psi $, in the near future.

\begin{table}
\begin{center}
\begin{tabular}{c|c|c|c|c}
$ J^{PC} (J^P) $ & Operator & Mass (MeV) & R/S & Candidate\\
\hline
\hline
$ 1^{--} $ &
$ \frac{1}{\sqrt{2}}[ (\ubar \gamma_5 \gamma_i \c)(\cbar \gamma_5 \u)
                   -(\cbar \gamma_5 \gamma_i \u)(\ubar \gamma_5 \c) ] $
        &   4238(31)(57)  &   R  & Y(4260) \\
$ 1^{--} $ &
$ \frac{1}{\sqrt{2}}[ (\sbar \gamma_5 \gamma_i \c)(\cbar \gamma_5 \s)
                   -(\cbar \gamma_5 \gamma_i \s)(\sbar \gamma_5 \c) ] $
        &  4405(31)(44)  & R  & \\
$ 1^{--} $ &$ (\cbar \gamma_i \c) (\cbar \c) $   &   6411(25)(43) & R \\
\hline
$ 1^{--} $ & 
$ \frac{1}{\sqrt{2}}\left\{[\u^T C\gamma_5\gamma_i \c][\ubar C\gamma_5\cbar^T]
  +[\u^T C \gamma_5 \c][\ubar C\gamma_5\gamma_i^T \cbar^T] \right\} $      &
4267(68)(83)  & R ? & Y(4260) \\
$ 1^{--} $ & 
$ \frac{1}{\sqrt{2}} \left\{[\s^T C\gamma_5\gamma_i\c][\sbar C\gamma_5\cbar^T]
  +[\s^T C \gamma_5 \c][\sbar C\gamma_5\gamma_i^T \cbar^T] \right\} $      &
4449(40)(55) & R  \\
\hline
\hline
$ 1^{++} $ & $\frac{1}{\sqrt{2}}[ (\ubar \gamma_i \c)(\cbar \gamma_5 \u)
                    -(\cbar \gamma_i \u)(\ubar \gamma_5 \c) ] $
        &   3895(27)(35) &   R  &  X(3872) \\
$ 1^{++} $ & $ \frac{1}{\sqrt{2}}[ (\sbar \gamma_i \c)(\cbar \gamma_5 \s)
                     -(\cbar \gamma_i \s)(\sbar \gamma_5 \c) ]  $
        &   4109(21)(32) &   R  &       \\
\hline
$ 1^{++} $ & $ \frac{1}{\sqrt{2}}\left\{(\u^T C\gamma_i \c)(\ubar C\gamma_5\cbar^T)
  -(\u^T C \gamma_5 \c)(\ubar C\gamma_i^T \cbar^T) \right\} $      &
 3891(17)(21)  & R &  X(3872)  \\
$ 1^{++} $ & $ \frac{1}{\sqrt{2}} \left\{ (\s^T C\gamma_i\c)(\sbar C\gamma_5\cbar^T)
  -(\s^T C \gamma_5 \c)(\sbar C\gamma_i^T \cbar^T) \right\} $      &
 4134(19)(25) & R &   \\
\hline
\hline
$ 1^+ $ & $\frac{1}{\sqrt{2}}[ (\ubar \gamma_i \c)(\cbar \gamma_5 \s)
                    -(\cbar \gamma_i \s)(\ubar \gamma_5 \c) ] $
        &   4007(34)(31) &   R  \\
$ 1^+ $ & $ \frac{1}{\sqrt{2}}\left\{(\u^T C\gamma_i \c)(\sbar C\gamma_5\cbar^T)
  -(\u^T C \gamma_5 \c)(\sbar C\gamma_i^T \cbar^T) \right\} $      &
 4015(25)(27) & R  \\
\hline
\end{tabular}
\caption{Masses of the lowest-lying states of the
molecular operators and the diquark-antidiquark operators. 
The column R/S denotes resonance (R) or scattering (S) states.
The entries with empty slots in the last column are theoretical predictions
\cite{Chiu:2005ey,Chiu:2006hd,Chiu:2006us}.}
\label{tab:mass_summary}
\end{center}
\end{table}

\section{Concluding remarks}

Note that we are working in the quenched approximation which
in principle is unphysical. However, our previous results on
charmed baryon masses \cite{Chiu:2005zc}, charmed meson masses, and 
our predictions of $ f_D $ and $ f_{D_s} $ \cite{Chiu:2005ue}
are all in good agreement with the experimental values  
\cite{Yao:2006px,Artuso:2005ym,Artuso:2006kz,Stone:2006cc}.
This seems to suggest that it is plausible to use the quenched lattice QCD 
with exact chiral symmetry to investigate the mass spectra of the 
charmonium-like 4-quark meson operators, 
as a first step toward the unquenched calculations.
The systematic error due to quenching can only be determined
after we can repeat the same calculation with unquenched
gauge configurations. (Note that Monte Carlo simulations
of unquenched gauge configurations in lattice QCD with exact
chiral symmetry, on the $ 20^3 \times 40 $ and 
$ 24^3 \times 48 $ lattices at $ \beta = 6.1 $, 
still remains a challenge to the lattice community.)
However, we suspect that the quenched approximation
only results a few percent systematic error in the mass spectrum,
especially for charmed mesons. Apparently, to gauge how well 
the quenched approximation in determining the mass spectrum of 
4-quark meson operators is to see whether the predicted states 
in Table \ref{tab:mass_summary} (entries with empty slots in the last column)
will be observed in high energy experiments. 
If it turns out that high energy experiments
do not find any of these predicted states, 
then it might imply that: (i) the mass spectra of 4-quark mesons 
in quenched QCD could be dramatically different from those 
in unquenched QCD, or (ii) the constancy of the spectral weight for 
different volumes with the same lattice spacing 
(i.e., $ W_{V_1}/W_{V_2} \simeq 1 $) 
might not be sufficient to guarantee that it is a resonance,  
or both (i) and (ii). On the other hand, if the predicted states
are confirmed by high energy experiments (even if our predicted masses 
turn out to be $5\sim8$\% off the experimental values), 
it would be a remarkable success for (quenched) lattice QCD with exact chiral 
symmetry. Obviously, no matter what is the outcome, it will be interesting.

\section*{Acknowledgement}

This work was supported in part by the National Science Council,
Republic of China, under the Grant No. NSC94-2112-M002-016 (T.W.C.),
and Grant No. NSC94-2119-M239-001 (T.H.H.), and by
the National Center for High Performance Computation at Hsinchu,
and the Computer Center at National Taiwan University.

\end{document}